\documentclass[10pt,twocolumn]{article}

\usepackage{graphicx}
\usepackage{amsfonts}
\usepackage{amsmath}
\usepackage{amssymb}
\usepackage{mathrsfs} 
\usepackage{lipsum}
\usepackage{color}
\usepackage{siunitx}
\usepackage{authblk}
\usepackage[margin=2cm]{geometry}
\usepackage{mathtools, cuted}
\usepackage{hyperref}

\usepackage{bm}
\usepackage{amsmath}

\title{Cheap and versatile humidity regulator for environmentally controlled experiments}
\author[1]{Fran\c{c}ois Boulogne}
\affil[1]{Laboratoire de Physique des Solides, CNRS, Univ. Paris-Sud, Universit\'e Paris-Saclay, Orsay 91405, France}

\date{\today}
\begin{document}

\twocolumn[
    \begin{@twocolumnfalse}
        \maketitle
        \begin{abstract}
            The control of environmental conditions is crucial in many experimental work across scientific domains.
            In this technical note, we present how to realize a cheap humidity regulator based on a PID controller driven by an Arduino microcontroller.
            We argue our choices on the components and we show that the presented designs can serve as a basis to the reader for the realization of humidity regulators with specific requirements and experimental constraints.
        \end{abstract}
    \end{@twocolumnfalse}
]

%
%
\section{Introduction}


Controlling the relative humidity is crucial in many experimental works across scientific domains.
To convince ourself with a scientific search engine, a query on ``effect of the relative humidity on'' leads to thousands of papers.
The control of humidity is not limited to direct applications such as studies on the mechanisms involved in evaporation of aqueous solutions, and of natural or manufactured materials, but also in other research areas.
To cite a few, humidity plays a crucial role in surface adhesion and tribology, \textit{e.g.} \cite{Feiler2005} and a spectacular manifestation can be seen with granular materials where humidity provides cohesion to the grains \cite{Fraysse1999}.
In biological related topics, effects of humidity have been found, for instance, on the development and the behavior of animals such as insects \cite{Guarneri2012}.
Furthermore, the spreading of a blood droplet on a surface and the resulting pattern  depends on the fluid properties and the evaporating conditions, which is important in biomedical applications and also in forensic investigations \cite{Bou-Zeid2013}.

To recall the basic thermodynamical concepts, the relative humidity $R_H$, generally expressed as a percentage, is defined as the ratio of the partial pressure of water in the gas to the partial pressure that is reached when the gas is in equilibrium with pure liquid water.
The latter is referred as the saturated vapor pressure, which depends on the temperature and the gas pressure.
The relative humidity must not be confused with the absolute humidity that corresponds to the water concentration in the gas phase.
The absolute humidity can be constant while the relative humidity varies if temperature and/or pressure change.
When $R_H=100\%$, the dew point is reached.

Since the early of the 20th century, a common method to regulate the relative humidity is based on the thermodynamical equilibrium of a saturated aqueous solution with the atmosphere.
Solutes, such as salts, modify the chemical potential of solutions.
Therefore, the relative humidity above the solution-vapor interface depends on the nature of the solute and the temperature.
Large datasets are reported in the literature and the interested readers can refer, for example, to these references \cite{OBrien1948,Solomon1951,Winston1960,Young1967,Greenspan1977}.
Thus, the principle is simple: a saturated solution, \textit{i.e.} a brine if salt is used, is placed in a crystallizer inside the controlled environment and the environment is regulated once the saturated solution is in equilibrium with the atmosphere.

However, this technique presents several limitations and drawbacks.
First, the accessible relative humidities are not continuously available at a given temperature.
In case the relative humidity must be varied for different experimental tests, the saturated solution must be replaced, which implies handling operations.
Also, this requires the purchase of a large variety of purified chemicals, which can represent a non-negligible cost.
In addition, reaching the thermodynamic equilibrium in the controlled environment is generally slow even with an artificial air circulation as noted for instance by Young \cite{Young1967}.
The convergence of the relative humidity is therefore difficult to estimate.
Finally, performing a humidity variation in time within a single experiment is not realistic.

Commercial humidity regulator apparatus are commonly sold for glove boxes.
They are often expensive and cannot be necessarily adapted to different experimental configurations.
In this paper, we propose to design a cheap humidity regulator based on a PID controller.
Our design has already been used successfully in research studies on water evaporation \cite{Dollet2017,Boulogne2018a}  and on the drying of silica colloidal suspensions \cite{Boulogne2015,Boulogne2016a}.
The regulator can be placed for example on commercial glove boxes, on home-made containers, or on the box of a high precision scale by replacing one door with a plastic sheet on which pipes are connected.
Also, the \textit{Do It Yourself} nature of our approach makes the use of such regulator appropriate for education and practical works where instrumentation and science can be advantageously combined with low budget requirements.

The aim of this paper is not to provide a unique recipe to apply, but rather to show the principles, to demonstrate the feasibility, and to provide the keys for custom developments fitting specific requirements.
We present two hardware designs, either based on a solenoid valve or on relays, respectively referred as design A and design B thereafter.
The design A is the safest and the faster to assemble, which makes it particularly suitable for education.
The design B is also presented as it offers alternates for the production of damped air.

\section{Principle and components}

\paragraph{Principle}

\begin{figure}
    \includegraphics[width=\linewidth]{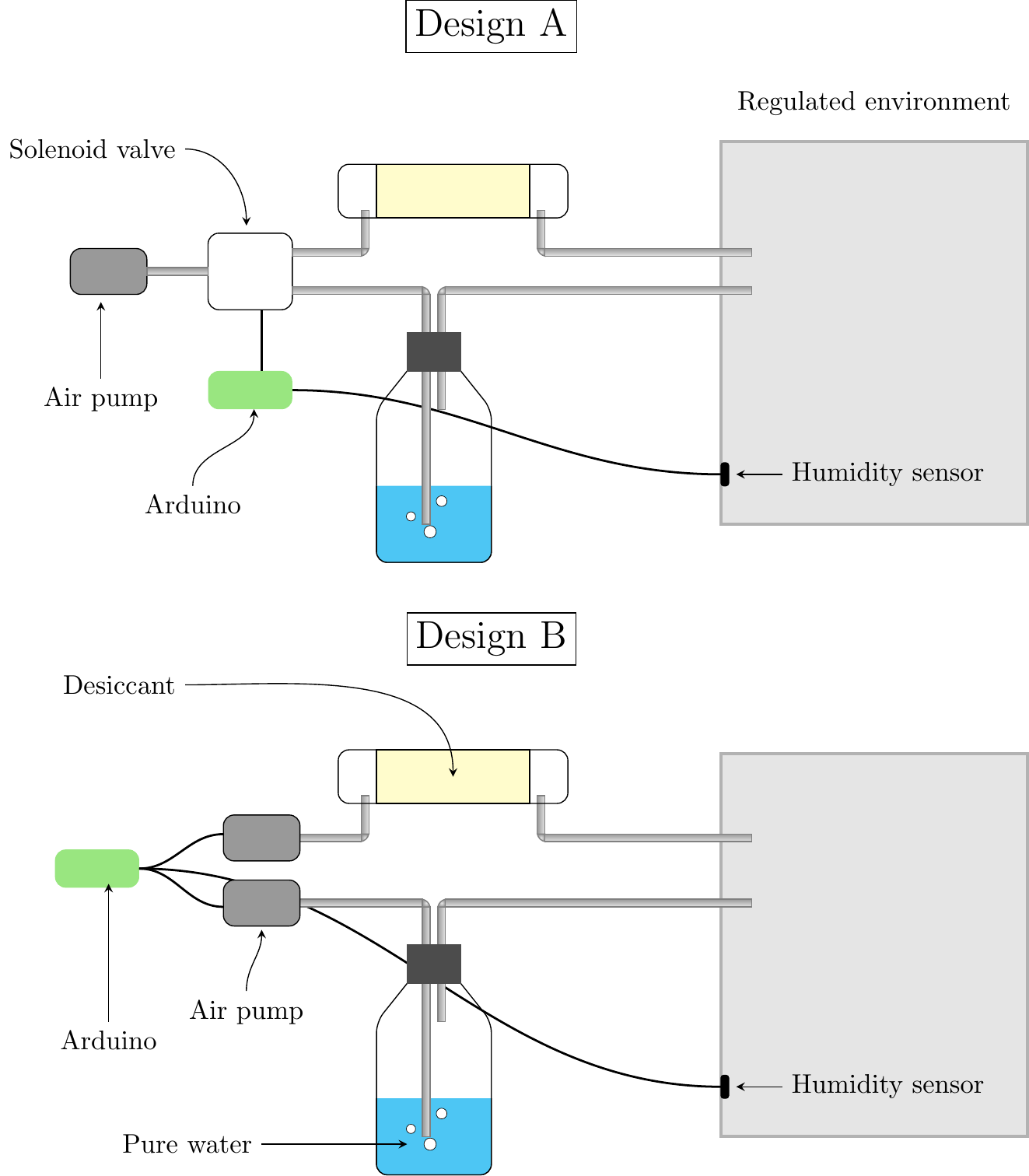}
    \caption{Principles of the two designs for humidity regulators.
    Design A (top) uses a single pump working continuously and connected to a two-way solenoid valve.
    Design B (bottom) is based on two air pumps working alternatively to inject dry or saturated air.
    }\label{fig:principle}
\end{figure}

The principle of the humidity regulator relies on a microcontroller that adjusts the relative humidity in the controlled environment.
The relative humidity is measured by a sensor and the microcontroller runs a proportional-integral-derivative controller (PID controller) to increase or decrease the water content in the atmosphere given a setpoint.
We present in this article two alternative designs presented in Fig.~\ref{fig:principle}.

\paragraph{Microcontroller}

The PID controller is made with an Arduino microcontroller, an open-source, cheap, accessible, and versatile device that can be reused for other purposes.
Among Arduino's boards, Arduino Uno is sufficient and we use the open source PID library\footnote{https://github.com/br3ttb/Arduino-PID-Library/}.

\paragraph{Relative humidity sensor}

Various technologies exist for relative humidity sensors, see \cite{Chen2005} for a review.
Here, we used an analog humidity sensor purchased from Honeywell, USA (model HIH-4021-003), a polymer based humidity sensor that can be obtained with an individual sensor calibration data.
Between 0 and $40^\circ$C, the sensor can be used in the full range of relative humidity.
The typical accuracy is $\pm 3.5\%$.
It is worth noting that this manufacturer provides sensors with or without protections against water condensation and also humidity sensors coupled with a temperature sensor.
As these sensors work with 5 VDC and have an analog output, they can be easily integrated to the Arduino Uno microcontroller.
The manufacturer indicates to use a 80 k$\Omega$ resistor.


\paragraph{Desiccant}

Air is dried by passing through a desiccant.
We use anhydrous calcium sulfate in a granular form purchased from Drierite, USA.
Once saturated, the desiccant can be regenerated by placing the material in a oven at a temperature around $230^\circ$C for two hours.
Some of these desiccants include a color indicator that shows the saturation of the calcium sulfate.
In case the indicator is useful, mixing colored and uncolored calcium sulfate reduces the total cost while keeping the visual advantage.
Commercial containers for desiccant referred as gas drying units exist but they can be easily replaced by plastic bottles pierced of holes to connect air pipes.

\paragraph{Damped air}
Damped air can be produced simply by bubbling air in pure water.
We made the attempt to use an air stone to decrease the bubble size and favor the vapor exchange, but we have not observed any improvement.

Bubbling air in water is preferred to ultrasonic humidifiers, which consists in vibrating a piezoelectric transducer to generate micron size droplets that evaporate rapidly.
Such approach has been used for instance by Gupta \textit{et al.} on their device \cite{Gupta2018}.
While this is interesting to increase rapidly the relative humidity, we found that finely controlling the quantity of water injected in the environment is particularly difficult and negatively impact the regulation quality.
Also the cloud generated by these devices can be problematic for visualization in the controlled environment.

\paragraph{Air pump}

For our controllers, we used aquarium air pumps, which are easily available with different flow capacities.
The flow rates typically span from 50 to 400 liters per hour and the choice is mainly motivated by the volume of air enclosed in the environment.
Typically, to control volumes of about 0.1 m$^3$, we used a Tetra APS 400 pump whereas for small environments such as the box of a high precision scale of about 3 dm$^3$, we chose the Tetra APS 50 version.
Air tubings sold for aquariums are perfectly appropriate as long as they are kink resistant.
To prevent a water backflow in the pump, we recommend to place the pumps at a higher level than the water bottle.

\paragraph{Solenoid valve and transistor}

For the design A, a solenoid valve is actuated to drive the air from the pump that works continuously.
The  three-way solenoid valve is purchased from Gems sensors (model MB332-VB33-L203), which has the advantage to allow for two paths of flow.
Therefore, a single valve is used while two two-way valves would have been necessary.
This solenoid valve is powered with a 12 VDC power supply.

The position of the valve is actuated by a TIP 120 NPN transistor.
The wiring to the microcroller needs a 1 k$\Omega$ resistor and a diode 1N4004.
The resistor limits the current drain and the diode prevents a reverse current when the valve is switched off.

\paragraph{Relays}

For the design B, relays are used to switch on/off the air pumps.
A relay is simply a switch driven by a voltage.
Relays are driven by the arduino board, so they must be 5 VDC, and must support the load voltage and current of the air pumps.
This setup requires two relays, one per pump.
Relays can be purchased directly soldered on a board  (SainSmart, USA) or assembled by the user.
For a better lifetime, solid state relays can be preferred to mechanical ones.
To wire the relays to the air pumps, we used extension cords for which the live wire (phase) is cut and connected on the two electrical terminals.
Naturally, the setup requires a particular caution as domestic electrical powers are manipulated.

\paragraph{Optional components}

Optional components can be added to the humidity controller.
For instance, we found convenient to have a manual control of the relative humidity setpoint by turning a knob and a direct reading of the measured relative humidity with a display.
As a result, the humidity controller is fully standalone and does not require a computer to change the setpoint.

To adjust the setpoint, we wired to the Arduino board a 10 k$\Omega$ potentiometer, for which the resistivity encodes the setpoint from 0 to 100\% divided in 1024 values.
For the screen, we choose a LCD display compatible with the Hitachi HD44780 driver for which the Arduino library \textit{LiquidCrystal} is designed.
In this paper, we do not show the wiring of this display for the sake of clarity of sketches.
Nevertheless, the instructions are available on arduino's website\footnote{\url{https://www.arduino.cc/en/Tutorial/HelloWorld}}.
Naturally, other types of display can be connected.


\begin{figure}[h!]
    \includegraphics[width=\linewidth]{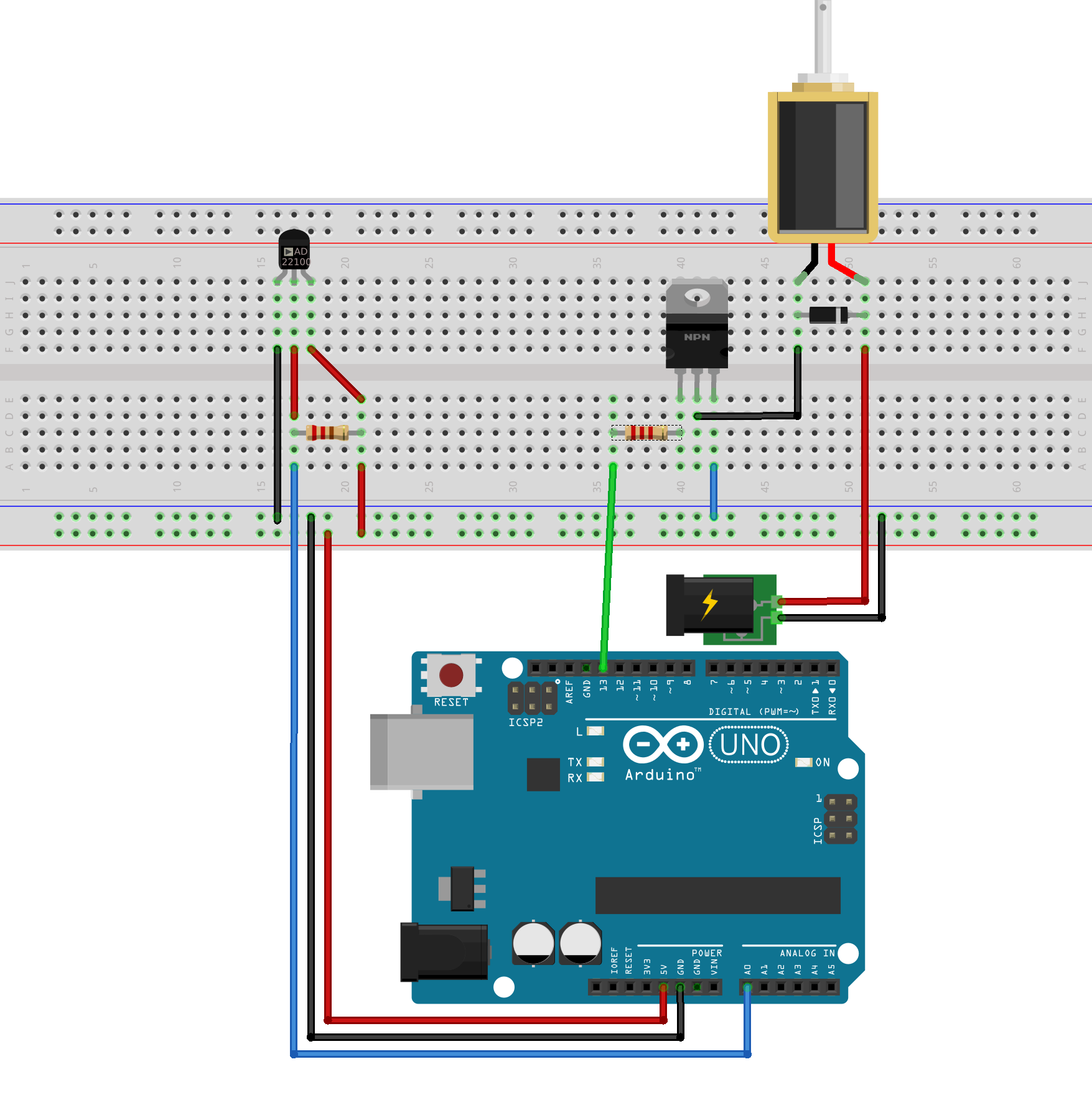}\\
    \includegraphics[width=\linewidth]{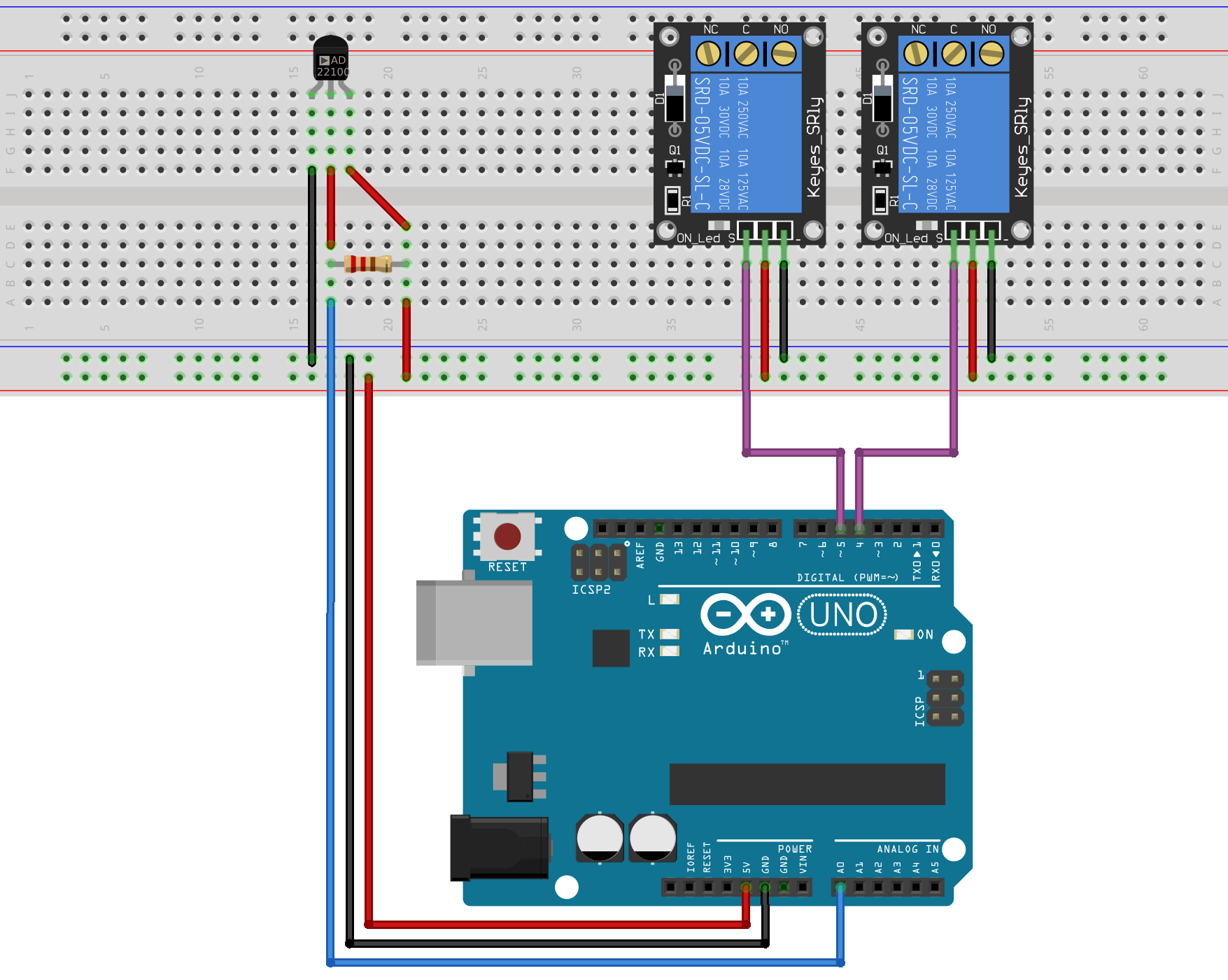}
    \caption{Top: design A relying on a solenoid valve actuated by a transistor.
    Bottom: design B based on two air pumps triggers by relays.
    The left parts of the sketches show the humidity sensor wiring.
    Optional components are not shown for the sake of clarity.
    }\label{fig:wiring}
\end{figure}

\section{Implementation}

\paragraph{Wiring and coding}
The setups are presented in Fig.~\ref{fig:wiring} for each designs.
Domestic electricity must be manipulated for the design B, which is risky without proper training and in the context of education.
Also, design A is generally faster to assemble as it requires less wiring.
Therefore, we encourage the reader to consider the design A.
Nevertheless, we wished to present design B for readers interested in using alternative ways to adjust the relative humidity, such as ultrasonic humidifiers.
The sketches presented in Fig.~\ref{fig:wiring} show the wiring for each design without the optional components.

The source codes are provided\footnote{\url{https://github.com/sciunto-org/humidity_regulator}} under BSD-3 license.
The repository includes codes for both designs, with and without optional components.
Adjustments in the source code are also necessary for the voltage-relative humidity conversion that is given on the calibration sheet of each sensor.
Another adjustment that the user must perform concerns the three PID constants, namely the proportional, the integral, and the derivative gains, which depend on air flow rates and box sizes.
The choice of these gains can be performed by following a systematic approach such as the Ziegler-Nichols method \cite{Ziegler1942}.
In practice, we tune manually these gains, first by increasing the proportional gain to get a sufficiently rapid variation of the relative humidity.
Then, the integral gain is increase to ensure the correct convergence and finally, the derivative gain is increased to limit the overshoot.

If the user wants to adjust the relative humidity setpoint without adding the optional components, this can be achieved through a serial communication by using a computer connected to the board.
The communication stream can be displayed for instance by the Arduino software with the serial monitors.

\paragraph{Costs}

\begin{table}[]
    \caption{Indicative prices in euros of the main components for each design in their minimal version.}\label{tab:costs}
    \begin{tabular}{l|l|l}
        & Design A & Design B \\\hline
        Arduino Uno    & 20       &  20      \\
        Tetra air pump &  25   &  2 $\times $25 \\
        Air tubing     & 4        &  4       \\
        Relay board    & --        &   9     \\
        Desiccant      & 30       &    30    \\
        Transistor     &   2     &    --     \\
        Solenoid valve &   50     &  --      \\
        12 VDC power supply & 10 & -- \\
        \hline
        Total          &    141   &     113  \\
    \end{tabular}

\end{table}

Cost estimates for both designs are given in Tab.~\ref{tab:costs}.
Some items such as wires, resistors, bottles are not listed as the cost is negligible or as they are part of the common lab supplies.
The total price is lower than 150 euros, which is particularly affordable compared to most of the commercial solutions.
Noteworthy, the cost of the 12 VDC power supply used for the solenoid valve can be easily avoided by recycling one from electronic waste.

\paragraph{Efficiency}

\begin{figure}[h!]
    \includegraphics[width=\linewidth]{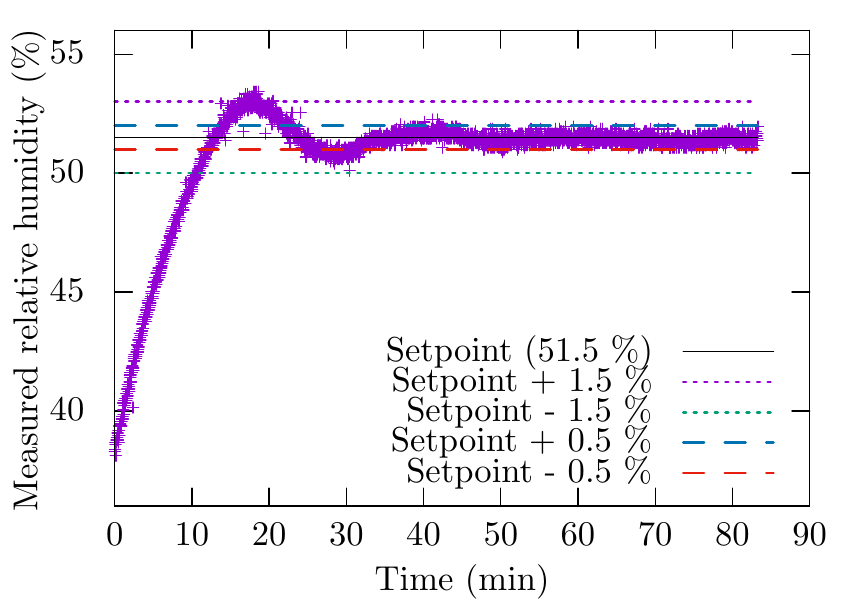}
    \caption{Measurement of the relative humidity in a box of $0.125$ m$^3$ during the regulation process for a setpoint at $51.5$ \%.}
    \label{fig:efficiency}
\end{figure}

Considering a regulated volume of $0.125$ m$^3$ and Tetra APS 400 air pumps configured with the design B.
Figure~\ref{fig:efficiency} shows the time evolution of the relative humidity in the box for a setpoint at $51.5$ \%  and an initial value at $38$ \%.
In these conditions, the setpoint is reached at $\pm 1.5$ \% within 10 minutes and is stabilized at $\pm 0.5$ \% within 40 minutes.
For equivalent conditions and air pumps, both designs provide an identical efficiency.

\section{Conclusion}

In this technical note, we presented the design of a cheap humidity controller based on
high quality and reliable components.
Beyond the cost argument, the presentation must serve as a starting point for education purposes and for custom designs that fits the variety of experimental requirements.
The \textit{Do It Yourself} approach enables quick additions: logging humidity values in a datafile, wire multiple humidity sensors to detect inhomogeneities, implement time variations of the setpoint, or change the setpoint from an event.

\section*{Acknowledgments}
The author is grateful to Alban Aubertin, Fran\c{c}ois Ingremeau, Jonas Miguet for stimulating discussions, Anniina Salonen for motivating the publication of this work, and Emmanuelle Rio for proofreading the manuscript.

\bibliography{biblio}

\bibliographystyle{unsrt}

\end{document}